\documentclass[A4,12pt]{article}
\usepackage{graphicx}%
\usepackage{multirow}%
\usepackage{amsmath,amssymb,amsfonts}%
\usepackage{amsthm}%
\usepackage{mathrsfs}%
\usepackage[title]{appendix}%
\usepackage{xcolor}%
\usepackage{textcomp}%
\usepackage{manyfoot}%
\usepackage{booktabs}%
\usepackage{algorithm}%
\usepackage{algorithmicx}%
\usepackage{algpseudocode}%
\usepackage{listings}
\usepackage{float}
\usepackage{graphicx}

\usepackage[margin=1in]{geometry}

\usepackage{natbib}
\usepackage{orcidlink}

\usepackage{natbib}
\def\BIC{\textsc{bic}}

\def\Mhat1{{\hat{M}_\omega^\dagger}}

\begin{document}





\author{B.W. Silverman\footnote{Emeritus Professor of Statistics, University of Oxford} \orcidlink{0000-0002-4059-2376}, 
\ L. Chan\footnote{Junior assistant professor, Università degli Studi del Piemonte Orientale ``Amedeo Avogadro"} \orcidlink{0000-0003-1193-6069},
\ K. Vincent\footnote{Independent Researcher and Consultant, kyle.shane.vincent@gmail.com} \orcidlink{0000-0002-3567-0798}, 
}



\title{Bootstrapping Multiple Systems Estimates to Account for Model Selection}
\maketitle

\abstract{
Multiple systems estimation using a Poisson loglinear model is a standard approach to quantifying hidden populations where data sources are based on lists of known cases. Information criteria are often used for selecting between the large number of possible models. Confidence intervals are often reported conditional on the model selected, providing an over-optimistic impression of estimation accuracy. A bootstrap approach is a natural way to account for the model selection. However, because the model selection step has to be carried out for every bootstrap replication, there may be a high or even prohibitive computational burden. We explore the merit of modifying the model selection procedure in the bootstrap to look only among a subset of models, chosen on the basis of their information criterion score on the original data. This provides large computational gains with little apparent effect on inference.  We also incorporate rigorous and economical ways of approaching issues of the existence of estimators when applying the method to sparse data tables.  
}

\section{Introduction}

Multiple systems estimation has been used as an approach for quantifying hidden populations of many different kinds.   It has been used specifically for human trafficking populations based on source data such as police records, non-governmental agencies, and outreach services. For an overall survey taking a view both of the range of applications and of methodology, see \cite{Bird2018}. For a recapitulation of a number of multiple systems estimation methods in policy contexts along with a survey of relevant data sets, see \cite{Silverman2019}.  
With administrative data sets becoming more publicly available over time, and data collection procedures reaching a wider and larger part of human trafficking populations, computationally stable and inexpensive multiple systems estimation procedures are desired for both researchers and practitioners.

Poisson loglinear regression \citep{Cormack1989} is one of the standard approaches. 
Recent developments include 
various approaches to model selection, and procedures tailored for sparse overlap in combinations of lists \citep{Cruyff2017, Chan2020}.

The log-linear method requires the choice of a model, specifying which terms are actually fitted to the data, and typically \citep{Baillargeon2007,Sil14} estimation is then conditional on the model selected for inference. 
Whatever approach is used in the frequentist paradigm, inference conditional on the selected model is likely to result in conservative standard errors and confidence intervals. 

A bootstrap approach, for example as set out in \citet[Section 3.3]{Chan2020}, makes it possible in principle to account for the model choice step in the inference procedure.
%
A stepwise model choice method may be  
sufficiently fast computationally to make the bootstrap feasible.  However, a longer established approach to model choice \citep{Baillargeon2007, Bales2015} is to choose among all possible models using the Bayes information criterion (BIC) and the first aim of this paper is to extend the bootstrap methodology to this method. The focus on the BIC is intended as illustrative not prescriptive; a topic for future work would be a comparison with other model choice criteria in the bootstrap context. 

If there are several lists, then the number of possible models is large and, as the BIC-based approach considers every possible model, the bootstrap approach leads to a computationally expensive and time-consuming procedure. 
In response to this limitation, we develop a modified bootstrap procedure that, for each bootstrap replication, only considers models that have high-ranking BIC scores on the original data.  We also give some consideration to other notions of closeness to the best-fitting model.  While we have only considered data sets which do not have concomitant information such as time dependence, the basic principles of our approach can be extended if such information is available and one wishes to choose between a large number of models.  It is unfortunate that there are, to our knowledge, no publicly available data sets of that kind. 

The current paper has been written from a frequentist viewpoint to encourage model selection to be taken into account when assessing estimation accuracy. There are several Bayesian approaches which, of their nature, avoid a focus on a single selected model. \cite{Kin:Bro2001} pioneered the use of a Markov Chain Monte Carlo approach to model fitting and averaging in this context. A comparison of a number of frequentist and Bayesian methods, including an MCMC approach, is given by \cite{Silverman2019}. 

A second focus of this paper is on dealing rigorously and economically with issues that arise when some combinations of lists are not represented in the observed data. This sparseness of data is typical in the human trafficking context. This part of the work builds on the results of \cite{Fienberg2012, Fienberg2012a}, which deserve greater attention in this area.  

Detailed consideration is given to the UK Home Office data set on human trafficking victims \citep{HomeOffice2014}, a data set on human trafficking in the New Orleans area \citep{Bales2020}, a data set on death occurrences in the Kosovo conflict from the late 1990s \citep{Ball2002}, and a data set on sex trafficked women in Korea \citep{hrdag2018}. The results indicate that our modification of the bootstrap procedure provides significant computational savings without altering the inference substantively. The newly-developed bootstrap procedure is applied to multiple systems estimation data sets, but the broad approach has obvious utility and application to other contexts where a statistical procedure involves the choice between a large number of candidate models using some numerical criterion and a bootstrap approach is used to carry out inference taking into account the effect of model selection. 

The paper is outlined as follows. Section 2 details the Poisson loglinear model, and sets out and develops the proposed bootstrap approach. Section 3 considers  the construction and existence of estimators when the model is applied to sparse multiple systems estimation data sets. Section 4 presents results based on empirical applications. Section 5 explores further ideas for searching for optimal models within bootstrap resamples. Software to implement the methods of the paper and to reproduce the results is available in the most recent version of the SparseMSE R package \citep{SparseMSE}.

\section{The model and bootstrap}

This section presents the Poisson loglinear model, sets out how a bootstrap approach can take account of model selection, and develops the idea of substantially reducing the computational burden by restricting the models considered in the bootstrap step.  

\subsection{The Poisson loglinear model}
Suppose that there are $t$ lists on which observed cases may fall. The set of lists on which a case is present is called its capture history $\omega$, a subset of $\{1, \ldots , t\}$. Define the order of a capture history to be the order of this subset. 
Define the capture count $N_\omega$ to be the number of cases with capture history $\omega$. 

Let $N^{\rm obs}$ be the vector of the $2^t - 1$ observable capture counts $\{N_\omega: \omega \ne \emptyset\}$. 
Define $N_{\rm total} = \sum_{\omega \ne \emptyset} N_\omega$, the number of cases actually observed.  The count $N_\emptyset$, not included in  $N^{\rm obs}$, is the `dark figure' of cases that do not appear on any list.

A model is defined by specifying a collection $\Theta$ of capture histories, always including the capture history $\emptyset$ and the $t$ capture histories of order 1. The model has parameters $\alpha_\theta$ indexed by the capture histories in $\Theta$, and the capture counts are modelled as independent random variables with
\begin{equation} \label{eq:modeldef}
N_\omega \sim \mathrm{Poisson}(\mu_\omega); \ \
\log \mu_\omega = \sum_ {\theta \subseteq \omega, \theta \in \Theta} \alpha_\theta.
\end{equation}
Note that 
both the capture counts $N_\omega$ and the model parameters $\alpha_\theta$ are indexed by capture histories.
Wherever possible we use $\omega$ as a suffix when the capture history refers to a count and $\theta$ when it indexes a parameter.

The parameters can be estimated by a generalized linear model approach, yielding parameter estimates $\hat{\alpha}_\theta$ and hence expected capture count estimates $\hat{\mu}_\omega$. The dark figure $N_\emptyset$ has expected value $\exp \alpha_\emptyset$ and so an estimate of the total population size $M$ is $\exp \hat{\alpha}_\emptyset + \sum_{\omega \ne \emptyset} N_\omega$.

It is natural to restrict attention to hierarchical models defined to have the property that if $\theta$ is in $\Theta$ then so are all $\theta'$ that are subsets of $\theta$. This can still leave a large number of possible models.   For example, for five lists, there are 6893 possible hierarchical models, excluding the non-identifiable model including all capture histories.  In general, let $\mathcal{H}_{t,l}$ be the set of all hierarchical models for $t$ lists allowing parameters/interaction effects of order up to $l$. If $l$ is omitted it will be assumed that $l=t-1$.
\subsection{The Bayes information criterion}
\label{sec:BIC}

The Bayes information criterion or BIC \citep{Schwarz1978} is one of the approaches to model choice 
implemented by \citet{Baillargeon2007}, and is one of the standard methods for multiple systems estimation  \citep{Cruyff2017, Chan2020}.  For any particular model $\Theta$, the BIC value $\mbox{\BIC}_\Theta$ is defined as
\begin{eqnarray}
\mbox{\BIC}_\Theta &= &
|\Theta| \log N_{\rm total} \nonumber \\ 
&& + 2 \sum_{\omega \ne \emptyset} ( \hat{\mu}_\omega - N_\omega \log \hat{\mu}_\omega +\log N_\omega ! ).
\label{eq:bicdef}\end{eqnarray}
To apply the BIC method, for $t$ lists allowing for models of order up to $l$, the model in  $\mathcal{H}_{t,l}$ with smallest $\mbox{\BIC}_\Theta$ is chosen.

There is some ambiguity about the appropriate sample size to use in the definition.  \citet{Baillargeon2007} consider the sample size to be the {\em case sample size} $N_{\rm total}$, the total number of cases  observed.  On the other hand another possibility, regarding the table of capture counts as the data, is to use the {\em capture sample size} $2^t - 1$, the number of observable capture counts.  There appears to be no strong argument of principle to determine which of these should be used.  We use the case sample size because it has become standard in this field; however our general approach is equally applicable for the capture sample size, as well as for other scoring criteria for model choice. 

%

\subsection{The bootstrap}

The bootstrap is a natural approach to take account of model selection when assessing the accuracy of estimation. The whole estimation procedure, including the choice of model, is carried out on each bootstrap replication. We focus attention on the bias-corrected and accelerated method set out by \citet{Efron1986}.  This method is second-order accurate and is invariant under transformation of the parameter space. 

The multiple systems estimation context allows for various computational economies. 
The $B$ bootstrap replications are obtained by simulating 
from a multinomial distribution with $N_{\rm total}$ trials and event probabilities proportional to the individual $N_\omega$.  For $i = 1, \ldots, B$, let $\hat{M}^{\rm boot}_i$ be the estimate of the total population obtained from the $i$th bootstrap replication, and let $\hat{M}$ be the estimate from the original data.


The bias-corrected and accelerated method makes use of two control parameters, the bias-correction parameter and the acceleration parameter.
The bias-correction parameter $\hat{z}_0$ is defined such that $\Phi(\hat{z}_0)$ is the proportion of the
$\hat{M}^{\rm boot}_i$ that are less than $\hat{M}$. 

Estimating the acceleration parameter $\hat{a}$ involves a jackknife step,
leaving out each of the original cases in turn and calculating an estimate of the population size from the resulting data. 
The following approach only requires the calculation of at most $2^t - 1$ estimates. 
Let $\Omega_1$ be the collection of all capture histories for which $N_\omega > 0$.   For each $\omega$ in $\Omega_1$, let $\hat{M}_{(\omega)}$ be the estimate of the population size from the original sample but with $N_\omega$ replaced by $N_\omega - 1$.  The number of estimates $\hat{M}_{(\omega)}$ is equal to $| \Omega_1 | \le 2^t -1$. Once these have been calculated, the average of the jackknife estimates is given by
$$
\hat{M}_{(\cdot)} = N_{\rm total}^{-1} \sum_{\omega \in \Omega_1} N_\omega \hat{M}_{(\omega)}
$$
and the acceleration factor by
 $$  \hat{a} = (1/6) S_2^{-3/2} S_3\
  $$
  where
  $$
   S_k = \sum_{\omega \in \Omega_1} N_\omega ( \hat{M}_{(\cdot)} - \hat{M}_{(\omega)})^k.
$$

As set out by \citet{Efron1986}, the bias-corrected and accelerated upper end-point of a one-sided $\beta$-confidence interval is the $\tilde{\beta}$ quantile of the $\{ \hat{M}^{\rm boot}_i \}$, where
$$
\Phi^{-1}(\tilde{\beta}) =
\hat{z}_0  +
\{ \hat{z}_0 + \Phi^{-1}(\beta) \}
[ 1 - \hat{a} \{ \hat{z}_0 + \Phi^{-1}(\beta) \} ]^{-1}.
$$
Note that, as is the case for the nonparametric bootstrap generally, the bootstrap replications are drawn with replacement from the original data, and therefore only the capture histories observed in the original data can be represented in the bootstrap samples.  The possibility of drawing from a smoothed version of the original data when using the bootstrap was considered in some detail by \cite{SilvermanYoung87}.  While in some circumstances smoothing can yield an improvement in the accuracy of estimation of a particular property of the underlying distribution, it can also possibly make matters worse.  The details of any particular case are technical and somewhat obscure, and so it is safest to make the more conservative approach of sampling from the observed data only, though this is a possible topic for future research.

\subsection{Restricting the models}
\label{sec:ntop}
Unfortunately the bootstrap computation can become prohibitive.  If $t=5$ and $l=4$ then the model choice for each data set requires the solution of 6893 generalized linear model problems.  So if there are $B$ bootstrap replications, the bootstrap process would involve fitting $6893B$ models, and therefore is computationally burdensome even with modern computing speeds.

To reduce the computation to manageable proportions, we approximate the bootstrap by restricting to a smaller number of models.  Fix a value $n_{\rm top}$, for example $n_{\rm top}=50$.  Apply the Bayes information criterion approach to the original data, and let $\mathcal{H}_{t,l}[n_{\rm top}]$ be the set of $\Theta$ in $\mathcal{H}_{t,l}$ giving the $n_{\rm top}$ smallest values of $\mbox{\BIC}_\Theta$.  Then restrict the bootstrap and jackknife steps to the models in $\mathcal{H}_{t,l}[n_{\rm top}]$.  Each bootstrap and jackknife choice will only involve $n_{\rm top}$ models rather than the much larger number of all hierarchical models.  The motivation is that even though a bootstrap or jackknife replication may result in a different choice of model from the original data, it is intuitively unlikely that a model that scored extremely badly on the original data would  have the best score on the replication.

\subsection{Varying the models considered }
\label{sec:ntopvary}

In assessing the effect of choice of $n_{\rm top}$ we calculate the bootstrap confidence intervals based on a range of values of  $n_{\rm top}$.  This can be done without performing the full calculation for each value separately.

\begin{enumerate}
\item Fix the largest value $n_{\rm top}^{\rm high}$ to be considered.
\item For each bootstrap replication $i$ and for each model $j$ in $\mathcal{H}_{t,l}[n_{\rm top}^{\rm high}]$ find the population size estimate and the BIC.  This gives $B \times n_{\rm top}^{\rm high}$ arrays of population size estimates $\hat{M}_{ij}^\prime$ and of BIC values $\beta_{ij}$.
\item  For each bootstrap replication $i$, find the indices of the record values of the vector $-\beta_{ij}$ with $i$ fixed.  Working recursively,  let $j_0 =  n_{\rm top}^{\rm high} + 1$ and, as long as $j_k > 1$, define $j_{k+1} = {\rm{arg\ min}}_{j <  j_k} \beta_{ij}$.   This gives a decreasing sequence $j_k$.  Finally construct a sequence of estimates $\hat{M}_{ij}$ by setting
$\hat{M}_{ij} = \hat{M}_{ij_k}^\prime$ for all $j \in [j_k, j_{k-1})$.
\item The array $\hat{M}_{ij}$ gives the bootstrap estimates for each value of $n_{\rm top}$ and for each bootstrap replication $i$.  The same process can be carried out for the jackknife estimates.  The bootstrap confidence intervals for every value of $n_{\rm top}$ can then be found.
\end{enumerate}
The main computational burden is in the calculation of the individual  $\hat{M}_{ij}^\prime$ and $\beta_{ij}$, so this approach allows for the effect of all $n_{\rm top}$ up to a given value $ n_{\rm top}^{\rm high}$ to be calculated for little extra effort than for the single value $ n_{\rm top}^{\rm high}$.  The maximum possible value for $ n_{\rm top}^{\rm high}$ is $|\mathcal{H}_{t,l}|$, the total of all models considered.  Increasing $ n_{\rm top}^{\rm high}$ beyond this value will make no difference, so we can equally consider this to correspond to $ n_{\rm top}^{\rm high} = \infty$, and this is the natural value to use for a general investigation of the approach.

\section{Sparse data}

\subsection{Constructing estimators}
\label{sec:miee}

In the context of modern slavery, particularly, it is common to observe data sets that are sparse in the sense that some of the $N_\omega$ are zero.  For any capture history $\omega$, define $N^*_\omega = \sum_{\psi \supseteq \omega} N_\psi$, the number of cases observed in the intersection of the lists in $\omega$ whether or not in conjunction with other lists.
Define the parameter $\theta$ to be $N^{\rm obs}$-empty if $N^*_\theta = 0$.



\citet[Section 2.4]{Chan2020} set out an algorithm for sparse tables with models whose parameters have maximum order 2.  The following algorithm is the natural generalization to $N^{\rm obs}$-empty parameters of any order.

\begin{enumerate}
\item Initially define $\Omega^\dag$ to be the set of all non-null capture histories and $\Theta^\dag = \Theta$.
\item For each $\theta$ in $\Theta$ for which $N^*_{\theta} = 0$, record the maximum likelihood estimator of $\alpha_{\theta}$ as $-\infty$ and remove $\alpha_{\theta}$ from the set of parameters $\Theta^\dag$ yet to be estimated.
\item For each such $\theta$ also remove from $\Omega^\dag$ all $\omega$ for which $\omega \supseteq \theta$ (because $N^*_{\theta} = 0$ the corresponding $N_\omega$ will all be zero).
\item Use the standard generalized linear model approach to estimate the parameters with indices in $\Theta^\dag$ from the observed counts of the capture histories in $\Omega^\dag$.  The set $\Theta^\dag$ comprises all the parameters in the model that are not estimated to be $-\infty$.
\end{enumerate}


This approach allows the parameters to take values in the interval $[-\infty, \infty)$, extending the real line only at the negative end.  The value $+\infty$ would not correspond to a finite parameter in the Poisson distributions of the observations, but $-\infty$ does make sense because these are parameters in a loglinear model, where a negative infinite parameter will correspond to a value of zero in the expected value, and hence the value, of the observed variable.  Following \citet{Fienberg2012, Fienberg2012a} we use the term {\em extended maximum likelihood} for estimates where the value $-\infty$ is allowed. \citet{Far2021} provide additional information on determining parameter redundancy for a loglinear model along with a method for deriving the estimable parameters.

\subsection{Existence of estimates}
\label{subsec:fienrinaldoprops}

\citet{Fienberg2012} show that the existence of the extended maximum likelihood estimate can be checked by the following linear programming problem.
For $\theta \in \Theta^\dag$ and $\omega \in \Omega^\dag$ as obtained in the algorithm set out above, define $A_{ \omega \theta} = 1$ if $\theta \subseteq \omega$ and 0 otherwise.  Let $\nu$ be the vector of values $N^*_\theta$ for $\theta \in \Theta^\dag$.
Let $s^*$ be the maximum value of $s$
over all scalars $s$ and all real vectors $x = (x_\omega, \omega \in \Omega^\dag)$ satisfying the constraints
\begin{equation}\label{eq:lp}
A^T x =\nu \;\;\; \mbox { and } \;\;\;
x_\omega - s \geq 0 \mbox{ for all $\omega \in \Omega^\dag$}.
\end{equation}
A necessary and sufficient condition for an extended maximum likelihood estimate of the parameters to exist is that $s^*>0$. 
We will then say that the Fienberg--Rinaldo criterion is satisfied.  For any model where the criterion is not satisfied, we will define, by convention, $\mbox{\BIC}_\Theta = \infty$, and so that model will not be chosen.

\begin{table}[htbp]
    \centering
    \begin{tabular}{||cccc||c|c|c|c||}
         \hline
         A & B & C & D  & $n_1$ & $n_2$ & $n_3$ & $n_4$\\ \hline
x & & & & 13 & 13 & 13 & 13\\
 & x & & &  16 & 16 & 16 & 16\\
 &  & x &  & 12 & 12 & 12 & 12\\
 &  &  & x& 11 & 11 & 11 & 11\\  &&&&&&&\\

x & x &  &  & 3 & 3 & 3 & 0\\
 & x &  & x & 4 & 4 & 4 & 4\\&&&&&&&\\
x & x & x &  & 2 & 2 & 0 & 2\\
x &  & x & x & 1 & 0 & 0 & 0\\
 & x & x & x & 1 & 0 & 0 & 0\\ \hline
             \end{tabular}
    \caption{Four possible observed data tables for a four list scenario}
    \label{tab:counterexample}
\end{table}
The criterion can only fail if there are zeroes in the data table.  If there are no zeroes then setting $x_\omega = N_\omega$ for all $\omega$ will yield a feasible solution with $s = \min_{\omega \in \Omega^\dag} N_\omega > 0$ and so the criterion will automatically be satisfied. 

There is no obvious hierarchical relationship to the pattern of zeroes in the data that will lead to a model failing the criterion.   Consider the example set out in Table \ref{tab:counterexample}.
The data vector $n_2$ has two zeroes in positions where $n_1$ is non-zero, while $n_3$ and $n_4$ each have one more zero than $n_2$.  For the model $[123,14]$  (A*B*C + A*D in standard R generalized linear model notation) the Fienberg--Rinaldo criterion is satisfied for $n_1$ and $n_3$ but not for $n_2$ and $n_4$.  So while removing two cells from the support of $n_1$ to get $n_2$ means that the condition becomes violated, removing one more cell from the support may or may not restore the existence of the extended maximum likelihood estimate.

In conclusion, it seems advisable always to check the Fienberg--Rinaldo criterion if the data are sparse. We will discuss possible economies in this process in Section~\ref{sec:fienrinaldobootstrap}.



\subsection{Economizing the check}
\label{sec:fienrinaldobootstrap}
Even after restricting to a smaller number of models, both the jackknife and the bootstrap steps involve evaluating the estimate for each particular model for a large number of possible data vectors.  We show below that the existence of the extended maximum likelihood estimate depends only on the support of the observed capture counts and not on their actual values.  This leads to the following approach, which allows for the Fienberg--Rinaldo checks to be carried out economically.

\begin{enumerate}
\item Remove from consideration all models which fail the check for the original data.
\item For the jackknife, check the condition for replications where the support is different from the original data, that is where $N_\omega = 1.$   
\item For the bootstrap, find all the distinct supports among the bootstrap replications; this will typically be much lower than the number of bootstrap replications.  For each support, construct a data vector which takes the value 1 on the support and 0 otherwise, and check the condition for that data vector.  The result will hold for all bootstrap replications with that support. 
\end{enumerate}

\textbf{Theorem} The existence of the extended maximum likelihood estimate for a given model and data depends only on the support of the observed capture counts.

\textbf{Proof}: Suppose $N^{(1)}$ and $N^{(2)}$ are two arrays of observed capture counts with the same support. 
Given a model $\Theta$, 
$\Theta^\dagger$ and $\Omega^\dagger$ only depend on the support of the data. 

For the model $\Theta$ and the data $N^{(1)}$, suppose that the Fienberg--Rinaldo criterion is satisfied, with the vector $x^{1}$ satisfying (\ref{eq:lp}) with $\min x^{(1)} = s^{(1)} > 0.$
Now choose an integer $k$ such that $k N^{(2)}_\omega \ge N^{(1)}_\omega$ for all $\omega$, possible because the support of both vectors is the same.
For $\omega$ in $\Omega^\dagger$,  define $\tilde{x}_\omega = kN^{(2)}_\omega - N^{(1)}_\omega$.   Define $\tilde{s} = \min_{\omega \in \Omega^\dagger} \tilde{x}_\omega  \ge 0.$
Then, for each $\theta$ in $\Theta^\dagger$, $(A^T \tilde{x})_\theta = k N^{(2) \star}_\theta - N^{(1) \star}_\theta = k \nu^{(2)}_\omega - \nu^{(1)}_\omega.$

Now define $x^{(2)} = k^{-1}(\tilde{x} + x^{(1)}).$   Then $A^T x^{(2)} = \nu^{(2)}$, and for each $ \omega \in \Omega^\dagger$ we have $x^{(2)}_\omega \ge k^{-1}(\tilde{s} + s^{(1)}) >0.$   Hence, for the model $\Theta$ and data $N^{(2)}$,  the Fienberg--Rinaldo criterion is also satisfied.
By applying the converse argument exchanging the roles of $N^{(1)}$ and $N^{(2)}$, we conclude that the criterion is satisfied for either both or neither of
$N^{(1)}$ and $N^{(2)}$, completing the proof.   

\section{Empirical applications}

We apply our bootstrap procedure to four real data sets. All analyses are, unless otherwise stated, based on 1000 bootstrap realizations.

\subsection{UK data}
\label{sec:UKdat}

\begin{figure}[htbp]
	\centering
\vspace{-1mm}
\centering
\includegraphics[width=0.8\textwidth]{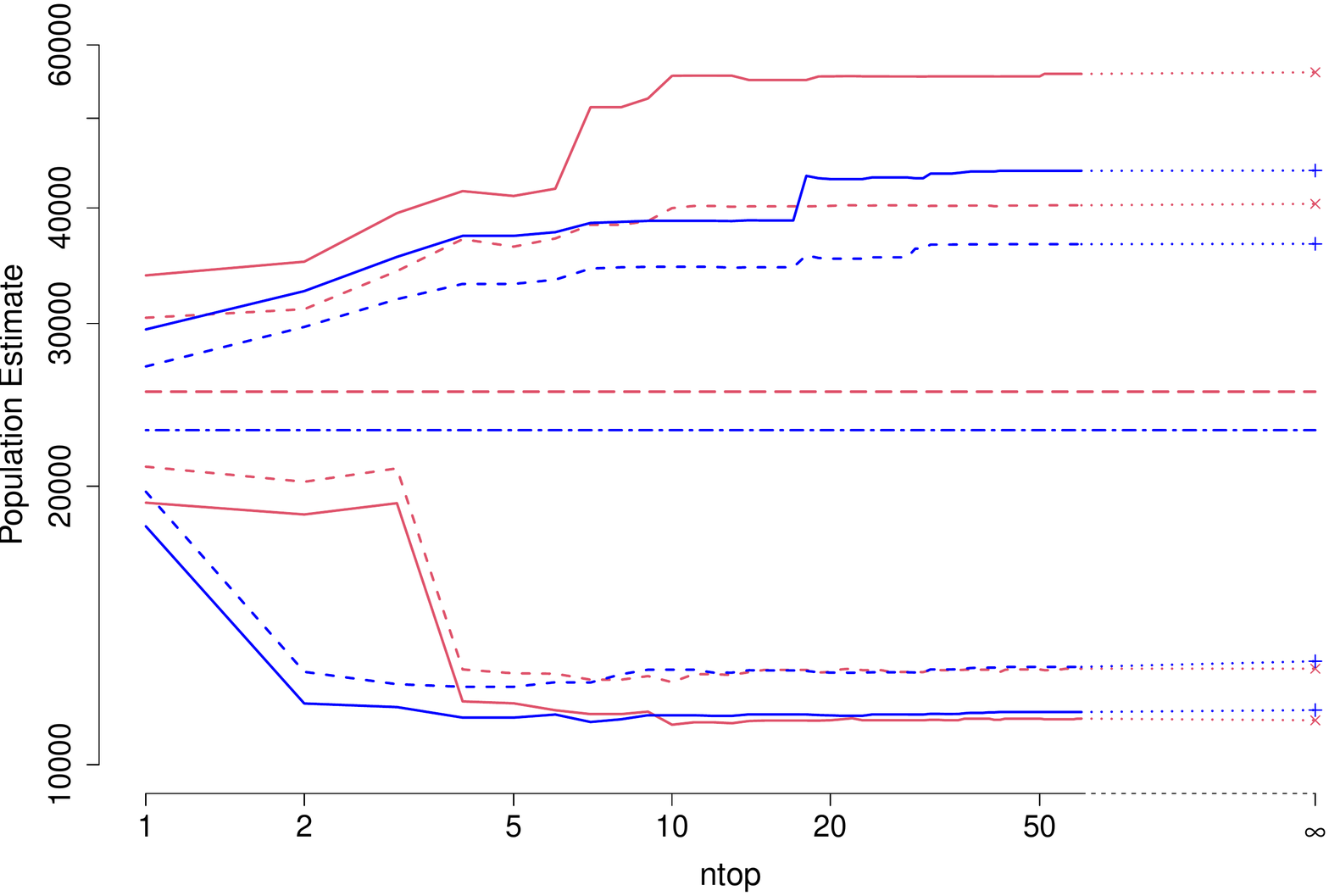}
\vspace{-3mm}
\caption{Estimates and confidence intervals for the five-list UK data.
Blue lines: maximum order 2; red lines: maximum order 4. Dot-dash horizontal lines: estimate; short dash lines: 80\% confidence interval; solid lines: 95\% confidence interval.}
\label{UKdat_5_maxorder_2and4}
\end{figure}

This five-list data set was studied in connection with the UK Modern Slavery Act 2015 \citep{HomeOffice2014} and discussed in detail in \cite{Bales2015}.  For other analyses and a fuller discussion, see for example  \cite{Silverman2019} and \cite{Chan2020}.
Figure \ref{UKdat_5_maxorder_2and4} shows the estimates and confidence bands for the estimated abundance based on varying choices for maximum order and $n_{\rm top}$. The value $n_{\rm top} =\infty$ is used to denote the full calculation where all possible models are considered, 1024 models in the case of maximum order 2 and 6893 for maximum order 4. The interval for $n_{\rm top}=1 $ is that obtained if we condition on the model chosen for the original data by the BIC criterion.

For both values of the maximum order, the results for $n_{\rm top}=60$ are virtually the same as those considering all models, while even $n_{\rm top}$ as small as 10 gives a result which is very close.   Using $n_{\rm top}=10$ will reduce the computation time for the bootstrap stage by a factor of more than 100 (considerably more for maximum order 4.)   Note also how much wider the confidence intervals are if allowance is made for the model choice.   As one would expect, the bands are somewhat wider if models of order 3 and 4 are allowed.  

\subsection{New Orleans data}
\label{sec:NewOrldat}

\begin{figure}[htbp]
	\centering
\vspace{-1mm}
\centering
\includegraphics[width=0.8\textwidth]{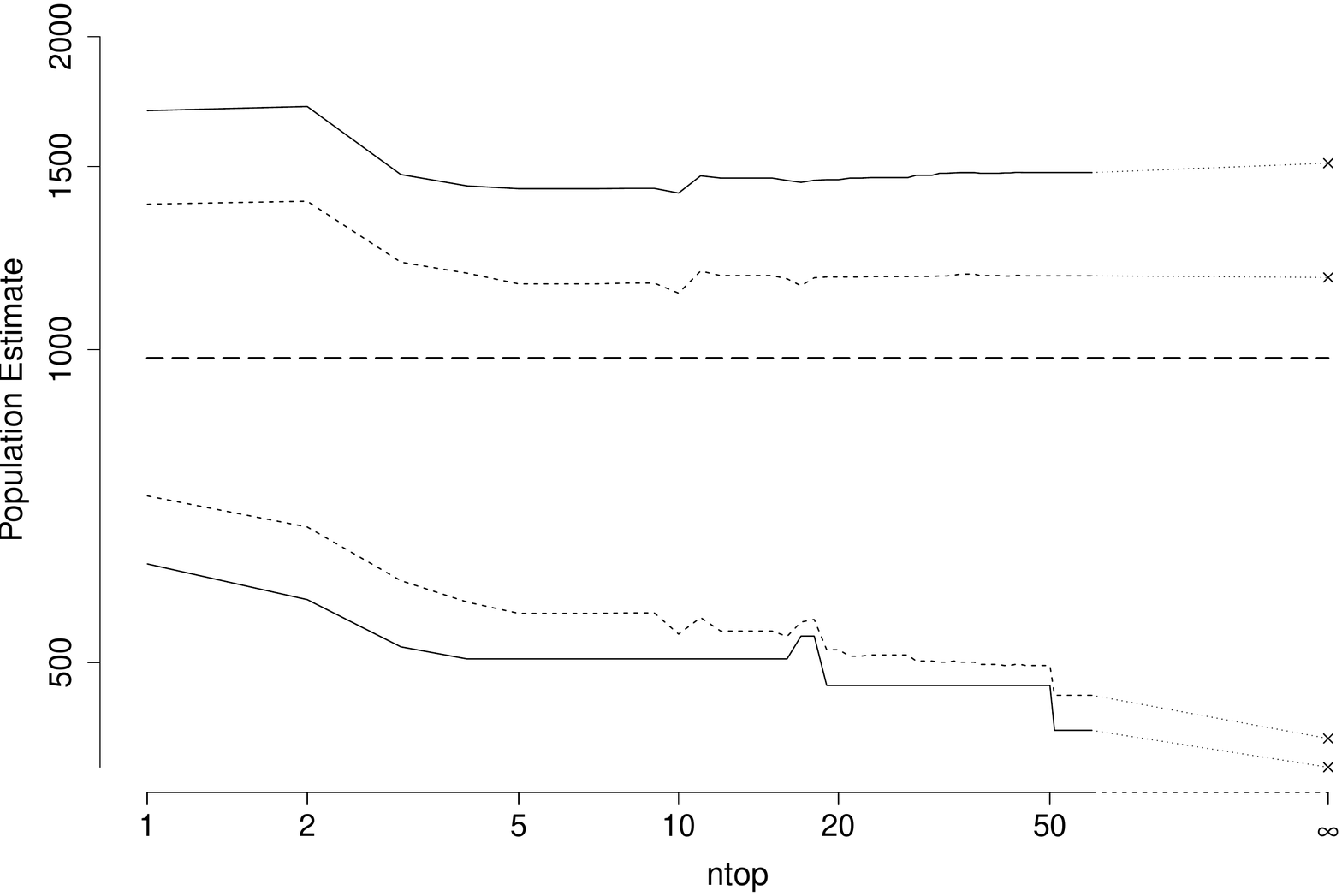}
\vspace{-3mm}
\caption{Estimates for five-list New Orleans data. 
Dot-dash horizontal lines: estimate; short dash lines: 80\% confidence interval; solid lines: 95\% confidence interval. The results for models of maximum orders 2 and 4 are essentially identical.}
\label{NewOrl_maxorder_2and4}
\end{figure}

This data set \citep{Bales2020} comprises 186 cases from New Orleans, distributed across 8 lists.  A five-list version is produced by consolidating the four smallest lists into one.  The results for maximum orders 2 and 4 are virtually identical, because the model with best BIC on the original data is of order 1, and all the other top models considered for $n_{\rm top} \le 60$ have order 2. All models with order 3 or 4 yield higher BIC values on the original data.   There is a very small difference in the upper 97.5\% confidence limit for $n_{\rm top}= \infty$, which is 1547 for maximum order 4 and 1544 for maximum order 2.  
Values of $n_{\rm top}$ greater than 20 give confidence intervals that are slightly narrow but not dramatically different from $n_{\rm top}= \infty$.  An $n_{\rm top}$ over about 50 makes the difference even less.

\subsection{Kosovo data}

Probably the first application  of multiple systems analysis in the human rights context was the estimation of the number of deaths in the Kosovo conflict in the late 1990s in connection with the trial of Slobodan Milosevich \citep{Ball2002,Ball2002a}.  From a total of 4400 cases collated between four lists, 
the method arrived at an estimate of 10,356 Kosovar Albanian deaths. The original analysis narrowed down to models which fitted the original data table well enough to yield a chi-squared statistic with a $p$-value of at least 0.05, but discarded those which fitted too well, placing an upper cutoff at $p=0.3$.  Among the models remaining, the choice was made by minimizing the chi-squared statistic divided by the residual degrees of freedom. In fact this model also minimizes the BIC. 

The bootstrap 95\% confidence interval conditioning on this model is [9100, 12000], rounding to the nearest 100; this accords closely with the profile likelihood 95\% confidence interval [9000, 12100] reported by \cite{Ball2002a}. It is instructive, to construct confidence intervals taking account of the model choice, and these are displayed in Table~\ref{tab:KosovoBCA}.  Even a value of $n_{\rm top}$ as small as 5 makes some allowance for the model choice; setting $n_{\rm top}=10$ gives the same confidence levels (within the rounding error) as if all the models were considered at the bootstrap stage, with more than a tenfold improvement in computational burden. 

\begin{table}[htb]
\centering
    \begin{tabular}{|l|l|l|}
\hline
$n_{\rm top}$& 80\% CI& 95\% CI \\ \hline
1               & [9500, 11300] & [9100, 12000]    \\
5               & [8500, 11500] & [8500, 17000]   \\
10              & [7400, 12200] & [6900, 18000]   \\
113 ($\infty$)  & [7400, 12200] & [6900, 18000]  \\ \hline
\end{tabular}
    \caption{Bootstrap confidence intervals for the Kosovo data, rounded to nearest 100 }
    \label{tab:KosovoBCA}
\end{table}

We also considered the possibility of applying the bootstrap to the original method of model choice.  Unfortunately, of the 1000 replications generated, 67 did not yield any models at all with chi-squared $p$-values in the specified range. With these replications excluded, the bootstrap 95\% confidence interval is $[7000, 15800]$; however because of the exclusion of 67 replications there is no clear justification for this result.  

\subsection{Korean data}
This data set was presented and analysed by \cite{hrdag2018} in connection with the quantification of Korean women forced into sexual exploitation in a particular location during the Japanese occupation of Indonesia in World War II. There are three lists, labelled B, C and D.  The total number of cases observed was 123, of which 12 occur on all three lists, 54 and 6 on the two-list intersections B$\cap$C and B$\cap$D, and 5, 5 and 41 on the single lists B, C and D respectively.  The authors' original analysis used a Bayesian nonparametric latent-class model approach \citep{Man16, Man17R} and provided a population estimate of 137, with a 95\% credible interval of (124, 198). 

\begin{table}[htb]
    \centering
    \begin{tabular}{|l|l|l|}
\hline
$n_{\rm top}$& 80\% CI& 95\% CI \\ \hline
1               & [136,  198] & [131, 248]    \\
2               & [135, 286] & [130, 348]   \\
6 ($\infty$)    & [135,  288] & [128, 349]  \\ \hline
\end{tabular}
    \caption{Bootstrap confidence intervals for various values of $n_{\rm top}$, for the Korea data}
    \label{tab:KoreaBCa}
\end{table}

Because there are three lists, there are eight possible models of order up to 2.  Two of these, $[12,13]$ and $[12,13,23]$, fail to satisfy the Fienberg--Rinaldo criterion on the original data and are removed from further consideration. The model $[12,23]$ has the lowest BIC score and yields a point estimate for the total population of 157.2, considerably higher than that produced by the method of \cite{hrdag2018}.  The 80\% and 95\% confidence intervals for various values of $n_{\rm top}$, rounded to the nearest integer and based on 1000 realizations, are as shown in the table. The confidence intervals are all shifted upwards relative to the credible interval found by \cite{hrdag2018}.

%

The result for $n_{\rm top}= 1$, conditioning on the best fitting model, gives considerably narrower confidence intervals than those taking  the model selection into account.  However, restricting to the two top models gives results virtually identical with those obtained by considering all six models for each realization.  There is no pressing computational need to restrict the number of models considered, but even in this case that number can be substantially reduced without materially affecting the result.

\subsection{Further considerations}
\begin{table}[htbp]
    \centering
    \setlength\tabcolsep{3pt}
    \begin{tabular}{|l|r r r r r |}
         \hline
         $n_{\rm top}$ & 1 &  5& 10& 50 & 100 \\ \hline
         Kosovo & 375 & 929 & 997 & 1000 & 1000 \\
         UK (maxorder 2)&      216 &  583 & 739 & 930 & 954 \\
         UK (maxorder 4)&   156 & 622  & 732 & 898& 943 \\
         New Orl (maxorder 2) & 357 & 622& 698 & 915 & 958 \\
         New Orl (maxorder 4) & 357 & 621& 697 & 913 & 956 \\
         \hline
    \end{tabular}
    \caption{For each data set and for each given value of $n_{\rm top}$, the number of bootstrap replications (out of 1000) for which the minimum BIC
    value is obtained within the $n_{\rm top}$ best models on the BIC criterion for the original data.}
    \label{tab:ntopbiccomparison}
\end{table}

The bootstrap runs for $n_{\rm top}= \infty$ facilitate a closer examination of the intuitive assertion that the BIC values from bootstrap replications are close to those from the original data.  Let $\rho_i$ be the Spearman rank correlation coefficient between the BIC values of the original data and the $i$th bootstrap replication, considering all models where both estimates exist. 
The average of the $\rho_i$ is around 0.97 for the Korea, Kosovo and New Orleans data. For the UK data it is 0.86 or 0.88 for models of orders up to 2 and 4 respectively. 

Another approach is to count the number of bootstrap realizations for which the global minimum of the BIC is achieved within the $n_{\rm top}$
best models on the original data.  See Table \ref{tab:ntopbiccomparison}.  The best model for the original data is the best model for only 15\% to 38\% of the bootstrap replications.

\section{Other possible approaches}
\label{sec:neighbours}

\subsection{Other measures of closeness}
 \label{sec:ktop}
 The discussion so far has been based on the intuitive idea that the best model for a bootstrap replication was likely to be one that scored well on the original data.  In this section, we consider models that are close to optimal on the original data in a broader sense. 
 
 Define the distance function between two models $\delta(\Theta, \Theta^*)$ to be $| \Theta \,  \triangle \,  \Theta^*|$, the size of the symmetric difference between the two sets of parameters.
 Define $\Theta$ and $\Theta^*$ to be $k$-neighbours if $\delta(\Theta, \Theta^*) \le k$.  
For any given model $\Theta$, define the BIC rank $r_1(\Theta)$ to be the rank of the BIC of the fit with model $\Theta$ to the original data, so that the best fitting model has $r_1(\Theta) = 1$. For integers $k \ge 2$, define $r_k(\Theta)$, the BIC rank of degree $k$, to be the smallest BIC rank of any $(k-1)$-neighbour $\Theta^*$ of $\Theta$.  The BIC ranks of all degrees up to any given $K$ can be found recursively using the following algorithm.

\begin{enumerate}
    \item Find the BIC ranks $r_1(\Theta)$ of all models from the estimation process using the original data.
    \item For each $\Theta$, let ${\cal N}(\Theta)$ be the set of all 1-neighbours $\Theta^*$ of $\Theta$, including $\Theta$ itself.   
    \item  Find the BIC ranks of increasing degree recursively, by finding
   $$r_k(\Theta) = \min_{\Theta^* \in {\cal N}(\Theta)} r_{k-1}(\Theta^*) 
    \mbox{ for $k = 2, \ldots, K$}. $$
           \end{enumerate}
%

The BIC ranks can be used in various ways to explore notions of closeness to the best model. As before, the basic principle is to narrow down the class of models considered, and then only to construct the bootstrap estimates on that smaller set of models.  For example, order the models in order of their BIC rank of degree 2, splitting ties by reference to their original BIC rank, and restrict attention to the $n_{\rm top}$ models according to this ordering.  We refer to this as the ``degree 2" approach and the original method as ``degree 1". Because both approaches will put the best model for the original data at rank 1, they will give identical results for $n_{\rm top} = 1$, while for $n_{\rm top} = \infty$ both approaches will also give identical results.  

\begin{table}[htb]
    \centering
    \begin{tabular}{|c|c|c|}
         \hline
         $n_{\rm top}$& degree &  95\% CI \\
         \hline
         1  & 1,2  &  [9100, 12000] \\
        \hline
        5   & 1     &  [7900, 17400] \\
            & 2     &  [8000, 13500] \\
        \hline
        10  & 1     &  [6900, 18000] \\
            & 2     &   [8000, 17700] \\
        \hline
        113 ($\infty$) & 1,2 & [6900, 18000] \\
        \hline
    \end{tabular}
    \caption{Bootstrap confidence intervals for various values of $n_{\rm top}$, for the Kosovo data. The approach using the BIC values only is referred to as degree 1, while degree 2 uses the BIC ranks of degree 2, tie-breaking using the original BIC ranks.}
    \label{tab:KosovoBCAcomparison}
\end{table}

 Table~\ref{tab:KosovoBCAcomparison} gives results for the Kosovo data. For $n_{\rm top}=10$, the confidence intervals using only the BIC values are virtually identical with those obtained considering all 113 possible models; there is a very small difference in the upper limit which is subsumed in the rounding.  However this approximation is not achieved by the degree 2 method. 
\begin{figure}[htbp]
	\centering
\vspace{-1mm}
\includegraphics[width=0.47\textwidth]{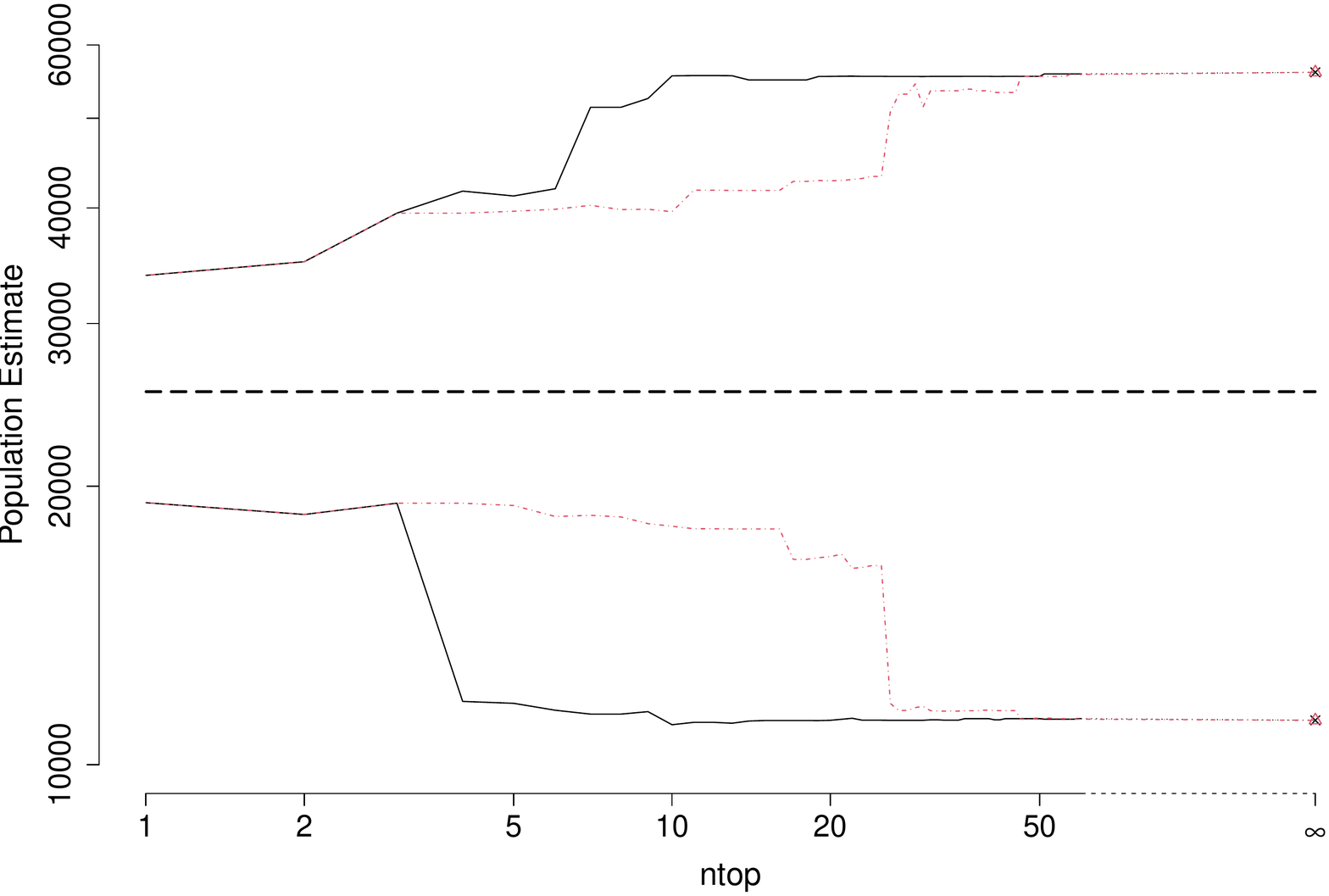}
\includegraphics[width=0.47\textwidth]{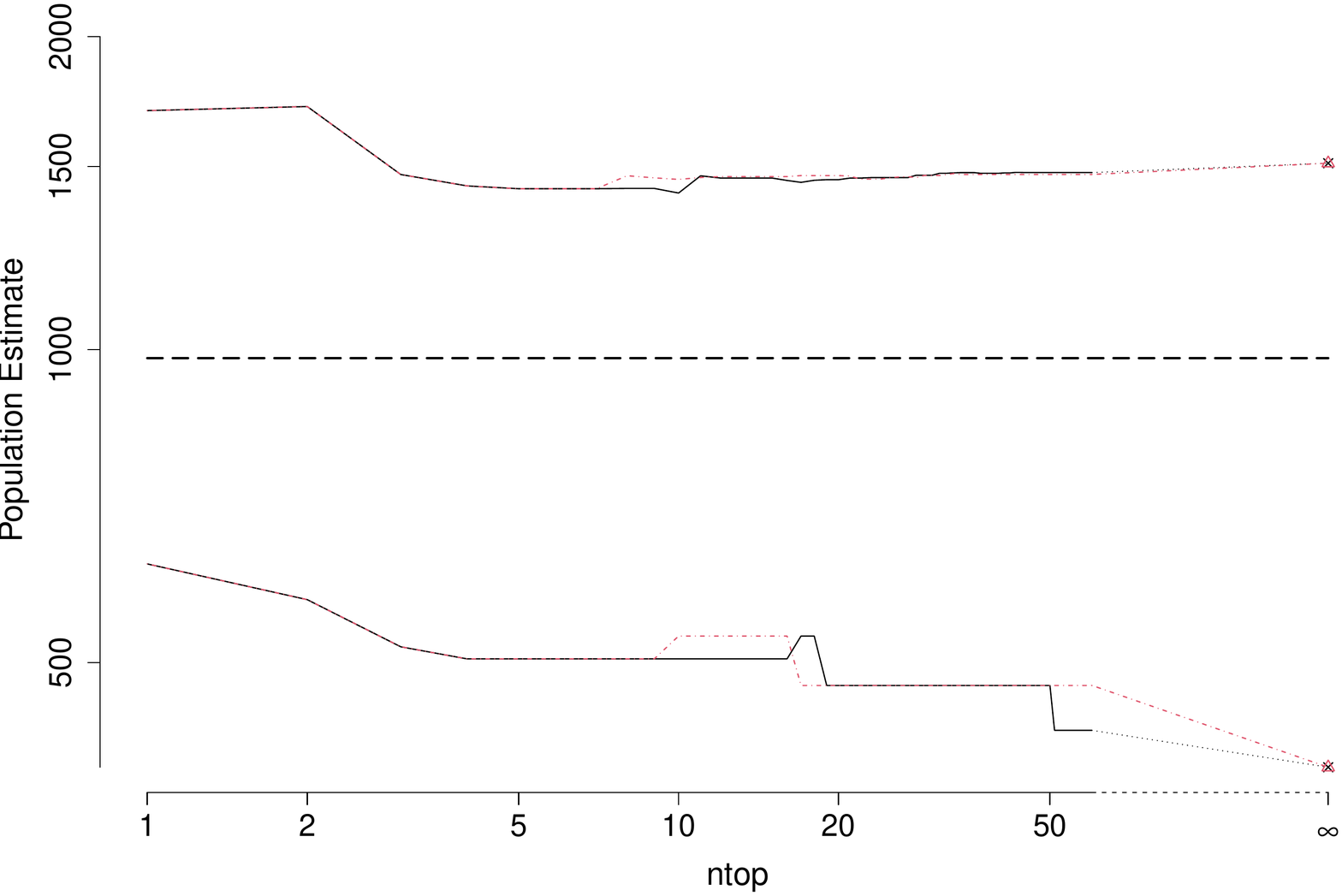}
\vspace{-3mm}
\caption{Estimate and 95\% confidence intervals for values of $n_{\rm top}$ up to 60, and $n_{\rm top} = \infty$, allowing models up to order 4.  Dashed line: estimate; black continuous: BIC values only used; red dotted: method using BIC ranks of degree 2.  Top: five-list UK data; bottom: five-list New Orleans data.}
\label{fig:UKandNOdegreecomp}
\end{figure}


 See Figure \ref{fig:UKandNOdegreecomp} for the UK and New Orleans data. For the UK data the degree 2 approach requires a larger value of $n_{\rm top}$  to obtain confidence intervals close to those for $n_{\rm top} = \infty$, though once $n_{\rm top} = 60$ there is little to choose.  For the New Orleans data the two approaches give virtually the same results. Restricting to models of order 2 has virtually no effect for either data set. 
 
\begin{figure}[htbp]
    \centering
    \includegraphics[width=0.8\textwidth]{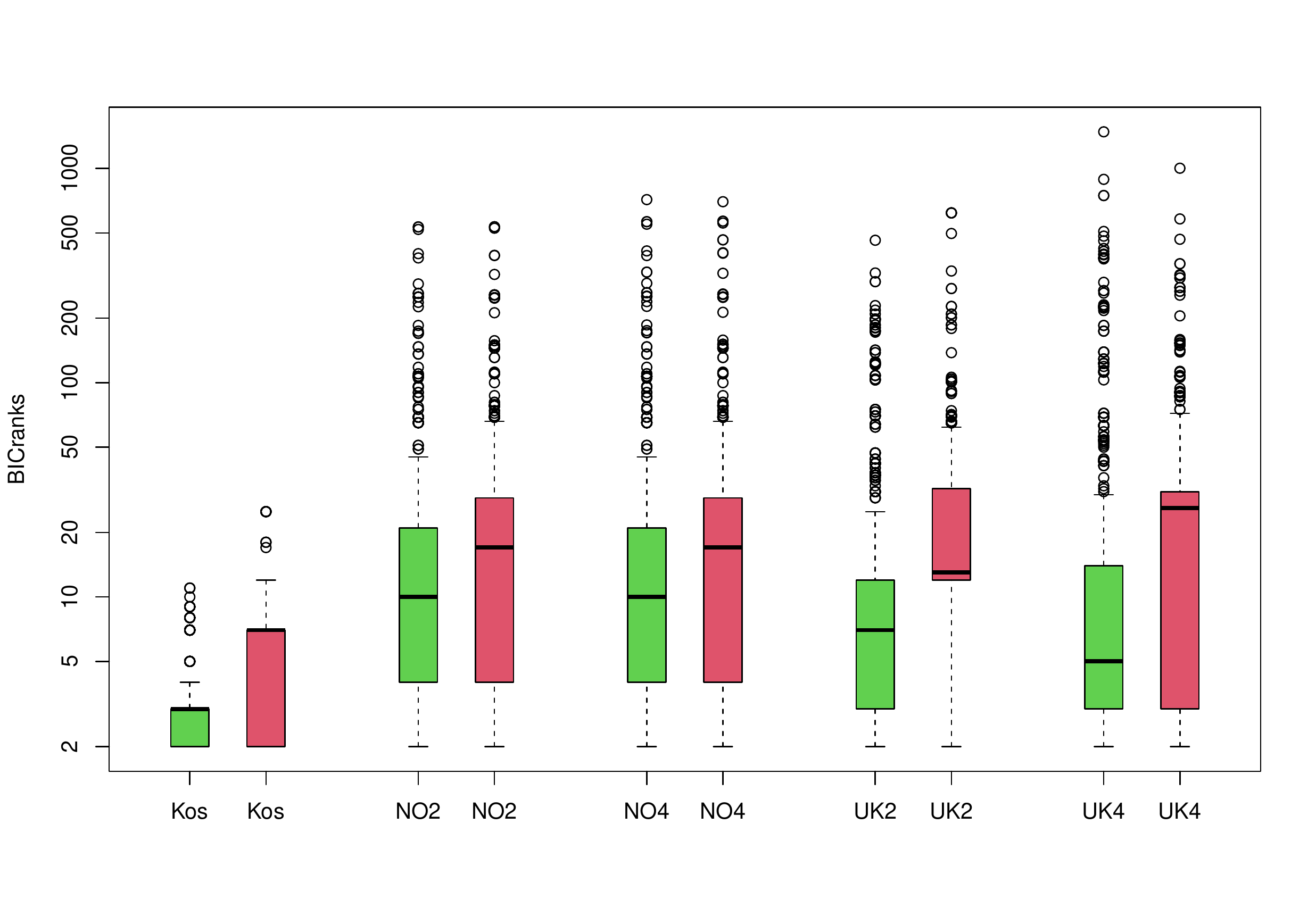}
    \caption{Boxplots of ranks (log scale) on original data of optimal BIC over bootstrap replications, omitting replications where the ranks are of value 1. Green plots: ranks of BIC itself; red plots: ranking using BIC ranks of degree 2.  Data sets/models, left to right: Kosovo; New Orleans with models of maximum order 2 and 4 respectively; UK maximum order 2 and 4 respectively.}
    \label{fig:BICranksboxplot}
\end{figure}

For another comparison, for each replication $j$, suppose that the model which minimizes the BIC over all models is at position $m^{(1)}_j$ for the degree 1 approach and $m^{(2)}_j$ for degree 2.  For any realization $j$ whose best BIC is observed for the best BIC model for the original data, $m^{(1)}_j = m^{(2)}_j = 1$.  Comparative box plots omitting such realizations are 
presented in Figure \ref{fig:BICranksboxplot}.   The rankings of the individual optimum BIC models are clearly lower for the degree 1 than the degree 2 method, a conclusion confirmed by Wilcoxon paired signed-rank tests.   The overall conclusion of this section is that considering neighbourhood information about models is an unwarranted complication.


\subsection{Greedy search for BIC minimum}
\label{subsec:downhillsearch}

An exhaustive search for the BIC minimum involves a large number of models, and  is infeasible for more than about five lists.  Even with four or five lists, bootstrap approaches require approximations. Furthermore, if there is a desire to conduct simulation studies then the computational burden becomes even more of an issue.

A possible alternative is a greedy search from one or more starting points to find at least a local minimum of the BIC or other criterion of interest.  The obvious starting point is the order one model with no higher-list terms.  At each stage all the neighbours of the current model are considered, obtained by either adding or removing a capture history from the parameter list, but only such that the hierarchical model property is preserved.  

\begin{table*}[htb]
\centering
    \begin{tabular}{|l|r|c|r|c|}
\hline
 & \multicolumn{2}{c|}{downhill search} & \multicolumn{2}{c|}{all models}\\
& \multicolumn{1}{c}{estimate} & \multicolumn{1}{c|}{95\% CI}
& \multicolumn{1}{c}{estimate} & \multicolumn{1}{c|}{95\% CI}
\\ \hline
UK5 all models   &12262             & [9330, 32115]  &25311  &[11168, 56052]\\
UK5 maxorder 2      &12262         & [9332, 28954]  &22991  &[11456, 43894]\\ \hline
UK all models  &12350 & [9622, 27889] &\multicolumn{2}{c|}{six lists}\\
UK maxorder 2 & 12350 & [9643, 25508] &\multicolumn{2}{c|}{} \\ \hline
NewOrl5  all models &981  & [397, 1564] & 981 & [397, 1511]\\ 
NewOrl5  maxorder 2 & 981  & [397, 1564] & 981 & [397, 1505]\\ \hline
NewOrl all models & 1110 & [586,  1512] &\multicolumn{2}{c|}{eight lists} \\
NewOrl maxorder 2 & 1110 & [586,  1512] &\multicolumn{2}{c|}{} \\ \hline
Kosovo all lists & 10357 & [7133, 18681] & 10357&[6947, 18010]\\
Kosovo maxorder 2 & 14342  & [12525, 16477] &14342 &[12451, 16273]\\ \hline
Korea & 157  & [128, 349] & 157 & [128, 349] 
\\
\hline
\end{tabular}
    \caption{Bootstrap 95\% confidence intervals for various datasets, comparing the downhill search approach with the consideration of all models. Numbers correct to the nearest integer.
    \label{tab:downhill_resvsntopinf_95_CI}}
\end{table*}

Because this method is quick to compute, it is practicable to repeat the estimation for each bootstrap replication.  There is no longer any need to restrict to only five lists, so for the UK and New Orleans data results were also found for the full data (six and eight lists respectively). Results are presented in Table 6.

For the UK five-list data the approach gives a lower estimate and narrower confidence intervals than those obtained from the consideration of all models.  To investigate further, we used the best minimum from five starting models of order 2, each containing five randomly chosen pairs of lists as parameters.  The same starts were used for the jackknife and bootstrap replications. The point estimates and confidence intervals are virtually the same as if all models are considered. A relatively small number of random starts for the five-list data thus exposes the existence of the second estimate whose BIC is lower than the original. \cite{Silverman2019} showed that, if a Bayesian approach is taken, the posterior for the UK five list data demonstrates an analogous bimodality.     

For the UK six-list data, it is no longer feasible to consider all models.  However as a check, downhill searches from random starts were also considered. Even using as many as fifty starting points never resulted in any estimate different from the one obtained from a single start from the null model. Although some downhill starts arrive at an estimate of around 21000, the BIC value is noticeably larger than that obtained by the search from the null model.  

The conclusion is that if downhill searches are used, it may be worth investigating random starts with the original data as a check; if no secondary minimum of the BIC is found, then it seems safe to proceed with downhill searching only from the null model.  As for the conclusion that should be drawn for the UK data, working with the six list data suggests a point estimate of around 12k with a confidence interval of 10k to 28k.  The estimate and lower confidence limit are similar to those obtained by a different model selection approach in the original paper \citep{Sil14}, though the upper limit is much higher.  From a policy point of view, this is reassuring, because it is widely acknowledged that the original estimate obtained in \cite{Sil14} is conservative.  

\section{Concluding remarks}
Multiple systems estimation can lead to the need to choose between a very large class of models.  The approaches we have explored show that it is possible, using appropriate approximations, to use computer-intensive approaches such as the bootstrap to take account of the model choice in making inferences from the data.  While we have concentrated on selecting a single model using BIC, our general approach is more widely applicable, for example if a different information or selection criterion is used, or if results are obtained by an ensemble of models weighted by reference to BIC or similar values.  In general, inference conditional on the selected model or models can lead to optimistic statements about the uncertainty of estimates, and it is hoped that the work presented here will help to facilitate more robust assessments of estimation uncertainty.




\bibliography{StatComprefs}

\begin{thebibliography}{}

\bibitem[Baillargeon and Rivest, 2007]{Baillargeon2007}
Baillargeon, S. and Rivest, L.-P. (2007).
\newblock Rcapture: loglinear models for capture-recapture in {R}.
\newblock {\em Journal of Statistical Software}, 19(5):1--31.

\bibitem[Bales et~al., 2015]{Bales2015}
Bales, K., Hesketh, O., and Silverman, B. (2015).
\newblock Modern slavery in the {UK}: How many victims?
\newblock {\em Significance}, 12(3):16--21.

\bibitem[Bales et~al., 2020]{Bales2020}
Bales, K., Murphy, L.~T., and Silverman, B.~W. (2020).
\newblock How many trafficked people are there in {G}reater {N}ew {O}rleans?
  {L}essons in measurement.
\newblock {\em Journal of Human Trafficking}, 6(4):375--387.

\bibitem[Ball and Asher, 2002]{Ball2002}
Ball, P. and Asher, J. (2002).
\newblock Statistics and {S}lobodan: Using data analysis and statistics in the
  war crimes trial of former president {M}ilosevic.
\newblock {\em CHANCE}, 15(4):17--24.

\bibitem[Ball et~al., 2002]{Ball2002a}
Ball, P., Betts, W., Scheuren, F., Dudukovic, J., and Asher, J. (2002).
\newblock {\em Killings and Refugee Flow in {K}osovo, {M}arch–{J}une, 1999: A
  Report to the {I}nternational {C}riminal {T}ribunal for the {F}ormer
  {Y}ugoslavia}.
\newblock American {A}ssociation for the {A}dvancement of {S}cience,
  Washington, DC.

\bibitem[Ball et~al., 2018]{hrdag2018}
Ball, P., Shin, E. H.-S., and Yang, H. (2018).
\newblock There may have been 14 undocumented {K}orean “comfort women” in
  {P}alembang, {I}ndonesia.
\newblock Report available at
  \url{https://hrdag.org/publications/there-may-have-been-14-undocumented-korean-comfort-women-in-palembang-indonesia/}.

\bibitem[Bird and King, 2018]{Bird2018}
Bird, S.~M. and King, R. (2018).
\newblock Multiple systems estimation (or capture-recapture estimation) to
  inform public policy.
\newblock {\em Annual Review of Statistics and Its Application}, 5(1):95--118.

\bibitem[Chan et~al., 2021]{Chan2020}
Chan, L., Silverman, B.~W., and Vincent, K. (2021).
\newblock Multiple systems estimation for sparse capture data: Inferential
  challenges when there are non-overlapping lists.
\newblock {\em Journal of the American Statistical Association},
  116(535):1297--1306.

\bibitem[Chan et~al., 2023]{SparseMSE}
Chan, L., Silverman, B.~W., and Vincent, K. (2023).
\newblock {\em Sparse{MSE}: Multiple systems estimation for sparse capture data
  in R}.
\newblock R package version 3.0.1.

\bibitem[Cormack, 1989]{Cormack1989}
Cormack, R.~M. (1989).
\newblock Log-linear models for capture-recapture.
\newblock {\em Biometrics}, 45(2):395--413.

\bibitem[Cruyff et~al., 2017]{Cruyff2017}
Cruyff, M., van Dijk, J., and van~der Heijden, P. G.~M. (2017).
\newblock The challenge of counting victims of human trafficking: Not on the
  record: A multiple systems estimation of the numbers of human trafficking
  victims in {T}he {N}etherlands in 2010-2015 by year, age, gender, and type of
  exploitation.
\newblock {\em CHANCE}, 30(3):41--49.

\bibitem[Efron and Tibshirani, 1986]{Efron1986}
Efron, B. and Tibshirani, R. (1986).
\newblock Bootstrap methods for standard errors, confidence intervals, and
  other measures of statistical accuracy.
\newblock {\em Statistical Science}, 1:54--75.

\bibitem[Far et~al., 2021]{Far2021}
Far, S.~S., Papathomas, M., and King, R. (2021).
\newblock Parameter redundancy and the existence of maximum likelihood
  estimates in log-linear models.
\newblock {\em Statistica Sinica}, 31(3):1125--1143.

\bibitem[Fienberg and Rinaldo, 2012a]{Fienberg2012}
Fienberg, S.~E. and Rinaldo, A. (2012a).
\newblock Maximum likelihood estimation in log-linear models.
\newblock {\em Ann. Statist.}, 40(2):996--1023.

\bibitem[Fienberg and Rinaldo, 2012b]{Fienberg2012a}
Fienberg, S.~E. and Rinaldo, A. (2012b).
\newblock Maximum likelihood estimation in log-linear models: supplementary
  material.
\newblock Available at
  \url{http://www.stat.cmu.edu/~arinaldo/Fienberg_Rinaldo_Supplementary_Material.pdf.}

\bibitem[{Home Office}, 2014]{HomeOffice2014}
{Home Office} (2014).
\newblock Modern {S}lavery {S}trategy.
\newblock Available at
  \url{https://www.gov.uk/government/publications/modern-slavery-strategy}.

\bibitem[King and Brooks, 2001]{Kin:Bro2001}
King, R. and Brooks, S.~P. (2001).
\newblock On the bayesian analysis of population size.
\newblock {\em Biometrika}, 88(2):317--336.

\bibitem[Manrique-Vallier, 2016]{Man16}
Manrique-Vallier, D. (2016).
\newblock Bayesian population size estimation using {D}irichlet process
  mixtures.
\newblock {\em Biometrics}, 72(4):1246--1254.

\bibitem[Manrique-Vallier, 2017]{Man17R}
Manrique-Vallier, D. (2017).
\newblock {\em LCMCR: Bayesian Non-Parametric Latent-Class Capture-Recapture}.
\newblock R package version 0.4.3.

\bibitem[Schwarz, 1978]{Schwarz1978}
Schwarz, G. (1978).
\newblock Estimating the dimension of a model.
\newblock {\em Annals of Statistics}, 6(2):461--464.

\bibitem[Silverman, 2014]{Sil14}
Silverman, B.~W. (2014).
\newblock {M}odern {S}lavery: an application of multiple systems estimation.
\newblock Published in conjunction with the UK Government Modern Slavery
  Strategy; available at
  \url{https://www.gov.uk/government/publications/modern-slavery-an-application-of-multiple-systems-estimation}.

\bibitem[Silverman, 2020]{Silverman2019}
Silverman, B.~W. (2020).
\newblock Model fitting in {M}ultiple {S}ystems {A}nalysis for the
  quantification of {M}odern {S}lavery: Classical and {B}ayesian approaches.
\newblock {\em Journal of the Royal Statistical Society: Series A},
  183:691--736.

\bibitem[Silverman and Young, 1987]{SilvermanYoung87}
Silverman, B.~W. and Young, G.~A. (1987).
\newblock The bootstrap: To smooth or not to smooth?
\newblock {\em Biometrika}, 74(3):469--479.

\end{thebibliography}

\end{document}